\documentstyle[12pt]{article}

\begin{document}
\begin{center}
{\large \bf 
A Simple Learning Algorithm for the Traveling Salesman Problem} \\
 
\vspace*{0.8cm}
 
{ \bf Kan Chen}\\
{ \sl Department of Computational Science}\\
{ \sl National University of Singapore, Singapore 119260}\\

\end{center}
 
\vspace*{0.8cm}
 
\begin{abstract}
We propose a learning algorithm for solving 
the traveling salesman problem based on a simple strategy of
trial and adaptation: i) A tour is selected by choosing
cities probabilistically according to the ``synaptic'' strengths between 
cities. ii) The ``synaptic'' strengths of the links that form the
tour are then enhanced (reduced) if the tour length is shorter (longer)
than the average result of the previous trials. We perform extensive 
simulations of the random distance traveling-salesman problem. 
For sufficiently slow learning rates, near optimal tours can be
obtained with the average optimal tour lengths close to the 
lower bounds for the shortest tour lengths.  
 
\end{abstract}

\pagebreak
Optimization problems are usually easy to formulate but hard to solve.
Particularly hard are a class of interesting optimization problems 
that are NP complete: an exact solution requires a number of 
computational steps that grows exponentially with the size of the problem.
The traveling salesman problem (TSP), which consists of finding the shortest 
closed tour connecting all cities in a map, is a classic example and a 
good testing ground for optimization methods. Because exact solutions are 
almost impossible to obtain the aim is to find near-optimal solutions.
 
A few optimization methods, based on ideas from physics and
biology, have been developed recently which
lead to rather good \underline{general purpose} algorithms 
\cite{boun} for solving optimization problems. They 
have been successfully  applied to a wide range of practical problems. 
One of them is a  stochastic algorithm 
known as optimization by simulated annealing (OSA) \cite{kir1}. 
The algorithm consists of an evolution according to Monte Carlo dynamics 
performed at a sequence of effective temperatures to simulate the 
annealing effect. The stochastic dynamics allows access to a larger region 
of configuration space than simple ``quenching'' methods,
and helps in reaching a good solution. However, the Monte Carlo search 
is based on the evolution of a \underline{single} configuration, thus is
often confined to a limited region of configuration space and
is not efficient in searching through the configuration space.
For this reason the performance of OSA is 
quite sensitive to the choice of an annealing schedule and the initial
configuration. Another general purpose algorithm which has been 
used extensively is the Genetic Algorithm (GA) \cite{holl}, 
which has also been applied to the traveling salesman problem \cite{brad}, 
\cite{muhlenbein}. The Genetic Algorithm demonstrates the importance of 
keeping many configurations (``species'') in the optimization process.
The algorithm mimics the evolutionary tools of reproduction,
mating, and mutation. It has been shown that the algorithm performs well 
for small-size TSPs \cite{brad}.
For large-size problems, however, it is not clear that current
genetic algorithms are efficient because one has to keep
a significant number of configurations. There are many variants 
of these two general algorithms using different local search methods.
For example, the annealing algorithm can be improved using
the multicanonical method; this  has also been applied to TSP \cite{lee}.
For the general performance of such algorithms, the reader is referred to
 the recent review article by Johnson and McGeoch \cite{johnson}.
 
In this paper, we propose a simple learning algorithm which
appears to have the advantages of both the OSA and Genetic algorithms. The 
preliminary version of our algorithm was presented in Ref.~\cite{chen1}.
Our algorithm may be viewed as a neural network algorithm, 
but differs completely from Hopfield and Tank's approach \cite{hopf},
 which is equivalent to gradient descent of an energy function with the 
tour length as a major term. In our approach the cities can be viewed 
as ``neurons'' and the connections between them as ``synapses''. 
Initially a synaptic strength $w_{ij}$ is assigned to each pair of cities 
\{ij\}: 
\begin{equation}
\label{init_weight}
      w_{ij}=e ^{-d_{ij}/T},
\end{equation}
where $d_{ij}$ is the distance between them and $T$ is a
parameter which controls the relative initial strengths of synapses.
The basic ingredients in the algorithm are
\underline{trial and adaptation}: first 
select a tour; then modify the synaptic
strengths of the links which formed the tour to favor tours with
shorter tour lengths by comparison. This procedure is then repeated. 

\underline{Selection of Tour}: A tour is selected 
by picking cities sequentially with the probability determined 
by the synaptic strengths. A simple selection procedure
consists of the following steps. The first city $i_{1}$ is picked at random, 
the second one $i_{2}$ is picked with probability $P_{i_{2}i_{1}} 
\propto w_{i_{2}i_{1}}$, then the third one $i_{3}$ is picked 
among available cities with 
probability $P_{i_{3}i_{2}} \propto w_{i_{3}i_{2}}$,
and so on. For computational efficiency, however, we set up   
 a priority list of neighbors for each cities and use the priority lists
 to generate the tour. 
The first two positions in the priority list are 
selected probabilistically with probabilities proportional to the 
corresponding synaptic strengths with the neighbors and the remaining 
positions in the list are ordered according to the synaptic strengths of 
the remaining neighbors. Since the synaptic strengths change slowly during 
the simulation, the new priority lists can be set up quickly before
each trial using  
the previous lists. The tour is now generated as follows. 
The first city $i_1$ is again picked at random, the second city $i_2$ 
chosen as the first available city in the priority list of $i_1$, 
the third city $i_3$ is chosen as the first available city in the 
priority list of $i_2$, and so on. 

The tour obtained is then improved using the exhaustive search for
two-bond re-arrangements \cite{lin}. This generates 
the tour of the current trial, which
will be used for comparison with the tours obtained 
from the previous trials. The search for two-bond re-arrangements 
allows tours of better quality, particularly in the beginning of the 
simulation process, to be used in tour evaluation and modification of 
synaptic strengths. This helps to improve the learning algorithm 
significantly. Note that with the use of the priority lists 
the exhaustive search can be made efficient, because only the links 
with low priority need to be swapped. As the learning process converges,
the configurations picked using the priority lists are already 
near-optimal, and very few two-bond re-arrangements are needed to 
improve them. 

\underline{Modification of Synaptic Strengths}:
The tour obtained in the current trial is compared
with the tours obtained in the previous trials. 
In our simulation we keep a number of tours (denoted by $m$ below) 
for the purpose of comparison . Let $\{i_1
i_2 ... i_N\}$ denote the current tour and $\{i'_1 i'_2 ... i'_N\}$ denote 
one of the previous tours. Let the tour lengths for these two tours
be $d$ and $d'$ respectively. The comparison of these two tours lead
to the following modification of the synaptic strengths of the links
that form tours:
\begin{eqnarray} 
       w_{i_l i_{l+1}}^{new}&=&w_{i_l i_{l+1}}^{old}
		 e^{-\frac{\alpha}{m} (d-d')}, l=1,...,N \nonumber \\
       w_{i'_l i'_{l+1}}^{new}&=&w_{i'_l i'_{l+1}}^{old}
		 e^{-\frac{\alpha}{m} (d'-d)}, l=1,...,N,
\end{eqnarray}
where $i_{N+1}=i_1$ and $i'_{N+1}=i'_1$. 
$\alpha$ represents the modification rate of synaptic strengths (since
$m$ comparisons are made at each stage, the coefficient in the exponent is 
written as $\alpha/m$). According to this rule, if the current tour 
length is shorter (longer) than the previous one the 
corresponding synaptic strengths for the links that form the tour
are enhanced (reduced). Another tour selection is made acording to the
prescription given earlier and the procedure repeated. As learning advances, 
some links are gradually abandoned, while others are increasingly favored: 
eventually the selection converges to a near-optimal tour.

In this paper, we study the random distance TSP, in which the distances
between cities are independent random variables between 0 and 1. Part of
the reason we choose the random distance TSP is that many algorithms 
based on simple local search methods
give very poor results for these instances. Besides, we can also
compare our results with the extensive numerical studies of
 Krauth and M\'{e}zard \cite{krauth}, who also obtained exactly
the length ($\approx 2.0415$) of the optimal tour in the limit of large 
number of cities for this problem.

Let us first discuss the choice of the parameters used in our simulation. 
The parameter $T$ (see Eqn.~\ref{init_weight}), is chosen to be
of the order of the shortest distance 
so that the synaptic strength for the long distance link is made very
small initially. In our simulation we simply choose 
$T=1/N$, where $N$ is the number of cities. The most important
parameter in our simulation is $\alpha$, which controls the rate of 
learning.  The choice of the parameter $m$ is not crucial. 
In our simulations we simply fix $m=50$ (i.e. we keep the 50 most
recent tours).

The general performance 
of our algorithm is illustrated in our study of a 200-city TSP. The 
result of the simulation with $\alpha=0.03$ is plotted in Fig.~1, 
which shows both the average tour lengths and the shortest tour length 
in successive intervals of 200 trials. As can be seen from the figure, 
the adaptive selection of tours  converges to a near-optimal one in the 
process of learning. At the beginning, the tour length
is comparable to that obtained by the exhaustive search for two-bond 
re-arrangements.  As learning advances, competition and ``mating''
of the tour segments in the tours used for comparison are effectively 
taking place through modification of synaptic strengths ---
this leads to tours with shorter and shorter tour lengths. The improvement
due to learning is quite drastic: the slow learning process
 reduces the tour lengths obtained at the begining of the 
learning process by as much as 40\%. The difference between the average
tour length and the shortest one in a given interval 
can be thought of as a measure of the effective ``temperature'', 
in analogy with OSA; during the course
of the simulation the effective temperature decreases as the learning
process converges. The dependence of the performance of the algorithm
 on the learning rate $\alpha$ is illustrated in Fig.~2, in which 
successive average tour lengths in the intervals of 200 trials are plotted 
for $\alpha=0.03, 0.12, 0.48$ and 1.92. As one might expect, 
better results are achieved for slower learning rates (small $\alpha$), but
with longer computer time. The longest optimazation
 with $N=400$ and $\alpha=0.03$ requires about 2 hours CPU on a 
SUNSparc workstation (Model 51).

To get a comprehensive picture of the performance of this algorithm and its 
dependence on the parameters, we have performed extensive simulations
for $N=25$, 50, 100, 200, and 400 with the number of samples equal to 800, 400,
200, 100, and 50 respectively. Large number of samples are needed for
small-size systems to obtain more reliable average optimal
tour lengths, because of the larger statistical fluctuations in optimal 
tour lengths for smaller sample sizes. In each simulation
both the shortest tour length and the convergence time, which is taken
to be the number of trials when the shortest tour length is first reached,
are recorded. Note that the actual CPU used is not proportional to the 
convergence time defined in this way, because
many two-bond re-arrangements are needed at the start of the
learning process and almost no two-bond re-arrangements are
needed at the end of the process. The average tour length for these
simulations are listed (together with the convergence time) in Table I 
and shown in Fig.~3.
 
We now compare our results with the known results.  
Krauth and M\'{e}zard \cite{krauth}, who used the Lagrangian 
one-tree relaxation of Held and Karp \cite{held} to obtain a lower bound
for the best tour length and the Lin-Kernighan algorithm \cite{lk} 
to obtain a upper bound. The performance of the Lin-Kernighan algorithm
depends somewhat on implementation. Better results are obtained in
the implementation presented in Ref.~\cite{johnson}.
The Lin-Kernighan algorithm is generally regarded
as one of the best algorithms for the TSP \cite{johnson} 
(in particular for the random distance TSP, where many other algorithms,
including simulated annealing, do not give good results). 
Our results are generally better than those 
obtained by the Lin-Kernighan algorithm implemented by
Krauth and M\'{e}zard (see Fig.~1. of Ref.~\cite{krauth})
 even with $\alpha=0.48$. It is comparable with
the Lin-Kernighan results of Ref.~\cite{johnson}, with $\alpha=0.03$. 
Note that our results for N=25 and 50 
using $\alpha=0.03$ and 0.12 are almost the same as the corresponding 
lower bounds, while our results for N=100, 200, and 400 using $\alpha=0.03$ 
are within 5\% of the corresponding lower bounds (see Fig.~1. of 
Ref.~\cite{krauth}). In contrast, the results of simulated annealing using 
two-bonds re-arrangement are 12\% and 37\% above the lower bounds for
$N=100$ and $N=316$ respectively \cite{johnson}. 
With smaller learning rate $\alpha$ we can get even better
results. This shows that a simple learning strategy can lead to a
very good optimization algorithm. However, like the other algorithms which 
involve stochastic searches, our algorithm is slow compared with the 
deterministic search algorithm like the Lin-Kernighan algorithm.

In conclusion, we have demonstrated how a complex optimization 
problem can be solved by a simple learning strategy of trial and
adaptation. The learning process is quite similar to the evolution
process in Genetic Algorithms. As in GA the algorithm has the
advantage that it is based on global
searches in configuration space, where configurations far apart 
in configuration space are searched.
 But instead of keeping many tour
configurations explicitly as in GAs we use 
``synaptic strengths'' to generate tour configurations 
probabilistically. Thus many tour configurations are implicitly kept for 
effective mutation and mating through the updating of the ``synaptic
strengths''. The version of the learning algorithm presented in this paper
for solving TSP may not be competive, but a better algorithm can be derived
using more sophisticated local search methods (such as 3-bond and 4-bond
moves for TSP). What we have shown is that the 
learning process based on gradual changes on ``synaptic strengths'' can
greately improve the results obtained by 
the corresponding local search methods. We believe that this learning 
strategy will be very valuable for other optimization problems,
in particular the ones where sophisticated local search algorithms
have not been found.

We thank C.\ Jayaprakash for critical reading of the manuscript and
for many helpful comments and suggestions. 

\pagebreak

\pagebreak   
 
\noindent Figure captions.
 
\vspace{6mm}

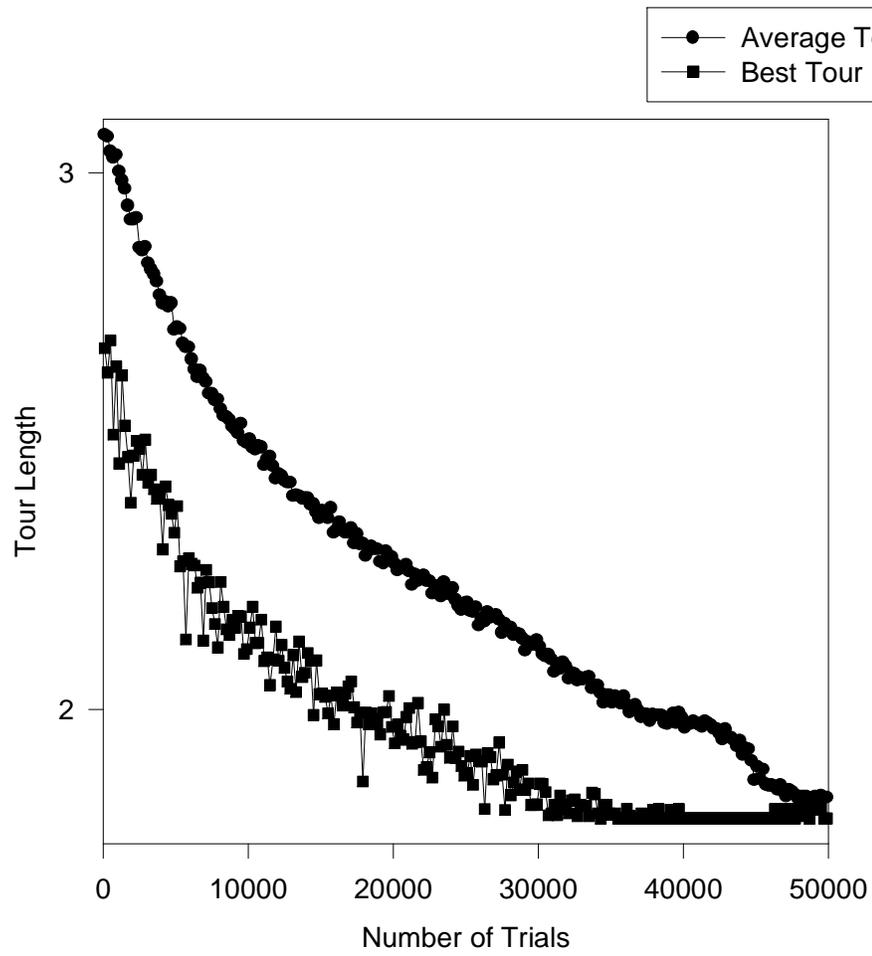
\noindent Figure 1. Tour length vs. number of trials for the
optimization of a 200-city random distance TSP: The upper curve 
shows the average tour lengths calculated in consecutive intervals of 
200 trials; the lower curve shows the shortest tour lengths in these 
intervals. Data are taken every 200 trials, and the optimization is 
performed with $\alpha=0.03$.

\vspace{6mm}

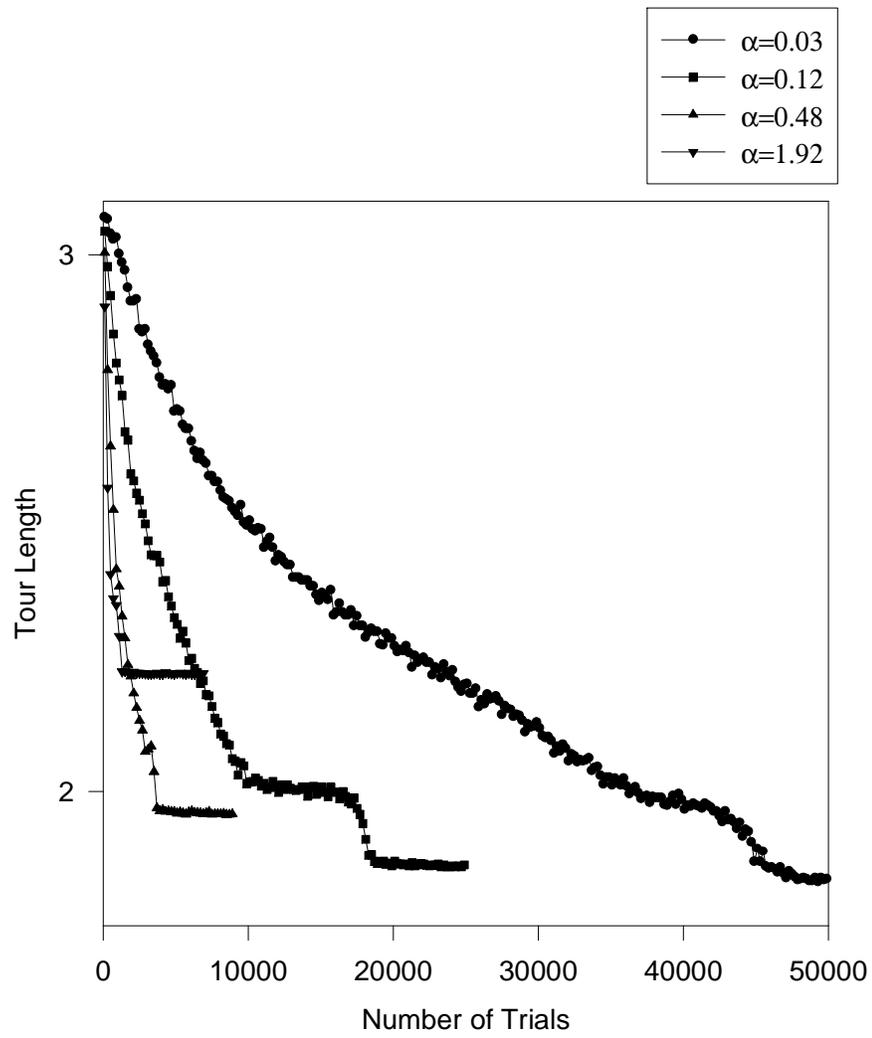
\noindent Figure 2. Average tour length vs. number of trials 
for the optimizations with $\alpha=0.03, 0.12, 0.48$ and 1.92. The
simulations use the same 200-city sample for which the
results shown in Fig.~1 are obtained.

\vspace{6mm}
 
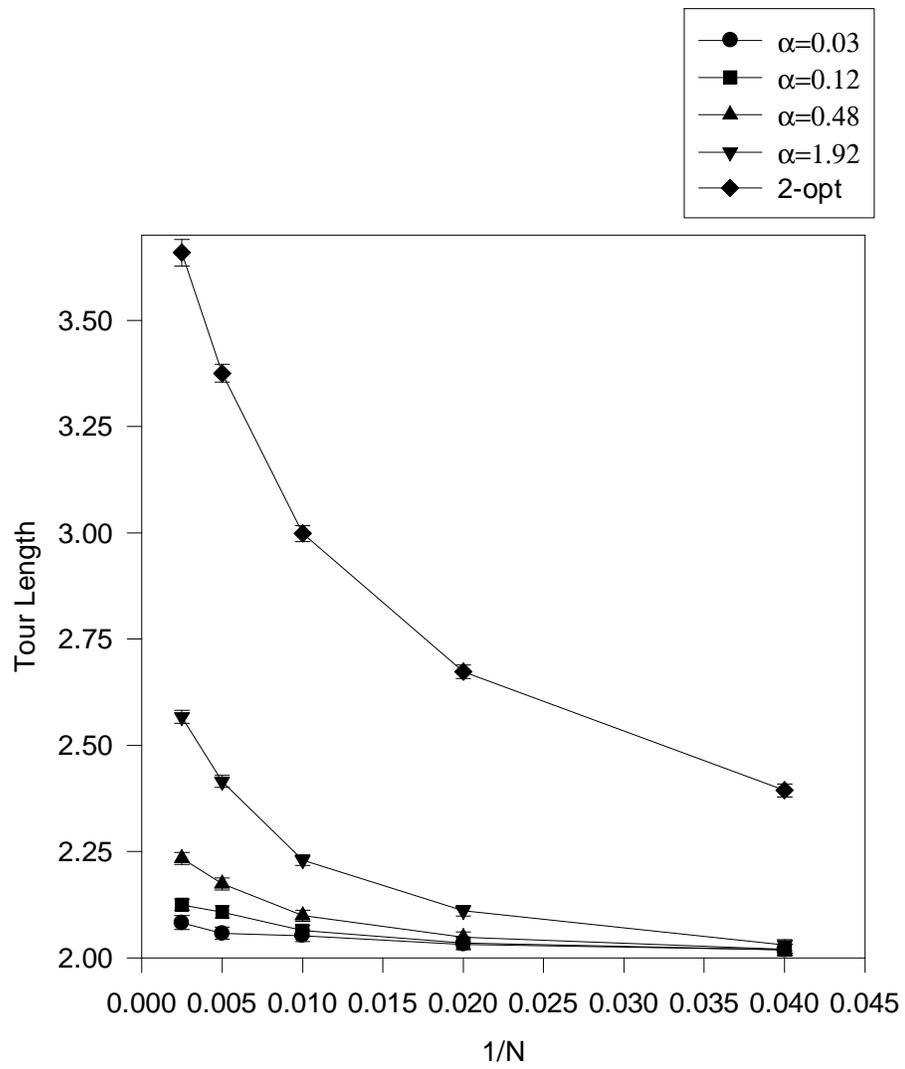
\noindent Figure 3. Optimal tour length for the random distance TSP.
The data on the top of the figure are obtained
using the iterative improvement method of exhaustive search for 
two-bond re-arrangements; below it, in order of decreasing tour lengths,
are a series of optimal tour lengths obtained 
using $\alpha=1.92,
0.48, 0.12$, and 0.03. Each data point represents an average over
800, 400, 200, 100, and 50 samples for N=25, 50, 100, 200, and 400 cities 
respectively. 
 
 \vspace{6mm}
 
\noindent Table captions.
 
\vspace{6mm}
 
\noindent Table I. Optimal tour length and convergence time
for a set of learning rate $\alpha$ and for the number of cities
 $N=25$, 50, 100, 200, and 400. For comparison, the results obtained using
the iterative improvement method of exhaustive search for two-bond
re-arragements (2-opt) are also listed in the table.

\pagebreak 
\begin{center}
Table I
\end{center}
 
\begin{tabular}{||c|c|c|c|c|c||} \hline
\multicolumn{6}{||c||}{Optimal tour length and convergence time} \\ \hline
    $N/\alpha$     &  0.03 & 0.12 & 0.48 & 1.92& 2-opt \\ \hline
25 &$2.019\pm	0.013$ & $2.019\pm 0.013$&	$2.020\pm 0.013$&
	$2.030\pm 0.013$& $2.394 \pm	0.015$ \\ \cline{2-6}
    & 1207 & 463 & 169 & 65 &  \\ \hline
50  &$2.032 \pm0.013$ & $2.035 \pm	0.013$ & $2.048 \pm 0.013$
	& $2.111 \pm 0.013$ & $2.673 \pm 0.016$ \\ \cline{2-6}
&7008 &2004 &571 &174 & \\ \hline
100 &$2.052 \pm 0.013$& $2.064 \pm 0.013 $&$2.099	 \pm 0.013$
&$2.230 \pm 0.013$ & $2.998	\pm 0.019$ \\ \cline{2-6}
&  19044 &   5205&   1466 & 418 &  \\ \hline
200 &$ 2.058 \pm 0.014$ &$ 2.108 \pm 0.014	$ &$ 2.174 \pm 0.014$ &
$ 2.415	 \pm 0.014$ & $3.375 \pm 0.021$ \\ \cline{2-6}
& 42683& 11731& 2948 & 894 &  \\ \hline
400 &$ 2.084 \pm 0.014$ &$2.125 \pm 0.013$ & $2.234\pm 0.014$
& $2.567 \pm 0.015$ & $3.659 \pm 0.030$  \\ \cline{2-6}
& 85775 & 21999 &  7107 & 1891 & \\       \hline
\end{tabular}
\end{document}